\documentclass[useAMS,usenatbib]{mn2e}
\usepackage{times}
\usepackage{rotating}
\usepackage{amsmath}
\usepackage{amssymb}
\usepackage[dvipsnames]{xcolor}
\usepackage{graphicx}
\usepackage{graphics}
\usepackage{mathrsfs}
\usepackage{hyperref}
\hypersetup{
     colorlinks   = true,
     citecolor    = blue
}

\def\apj{{ ApJ}}
\def\apjl{{ApJL}}

\def\mnras{{ MNRAS}}
\def\araa{{ ARA\&A}}

\def\nat {{ Nature}}

\def\prd{{ Phys. Rev. D}}

\def\pasa{{PASA}}

\makeatletter
\newcommand{\lambdabar}{{\mathchoice
  {\smash@bar\textfont\displaystyle{0.25}{1.2}\lambda}
  {\smash@bar\textfont\textstyle{0.25}{1.2}\lambda}
  {\smash@bar\scriptfont\scriptstyle{0.25}{1.2}\lambda}
  {\smash@bar\scriptscriptfont\scriptscriptstyle{0.25}{1.2}\lambda}
}}
\newcommand{\smash@bar}[4]{%
  \smash{\rlap{\raisebox{-#3\fontdimen5#10}{$\m@th#2\mkern#4mu\mathchar'26$}}}%
}
\makeatother

\long\def\symbolfootnote[#1]#2{\begingroup%
\def\thefootnote{\fnsymbol{footnote}}\footnote[#1]{#2}\endgroup}
\newcommand{\gae}{\lower 2pt \hbox{$\, \buildrel {\scriptstyle >}\over {\scriptstyle
\sim}\,$}}
\newcommand{\lae}{\lower 2pt \hbox{$\, \buildrel {\scriptstyle <}\over {\scriptstyle
\sim}\,$}}

\def\d{\mathrm{d}}

\def\mr{\mathrm}
\def\mc{\mathcal}
\def\pp{{\prime\prime}}

\def\r5{\rho_{_5}}

\def\para{\parallel}

\def\t{\widetilde}

\def\bphi{\beta_{\phi}}

\begin{document}

\title[FRB Polarization]
{Fast radio burst source properties from polarization measurements}


\author[Lu, Kumar \& Narayan]
  {Wenbin Lu$^{1,2}$\thanks{wenbinlu@tapir.caltech.edu},  Pawan
    Kumar$^2$\thanks{pk@astro.as.utexas.edu} and Ramesh
    Narayan$^3$\thanks{rnarayan@cfa.harvard.edu}\\ 
  $^1$TAPIR, Walter Burke Institute for Theoretical Physics, Mail Code
  350-17, Caltech, Pasadena, CA 91125, USA\\
$^2$Department of Astronomy, University of Texas at Austin, Austin,
TX 78712, USA\\
$^3$Harvard-Smithonian Center for Astrophysics, 60 Garden Street,
Cambridge, MA -2138, USA}


\maketitle

\begin{abstract}
Recent polarization measurements of fast radio bursts (FRBs)
provide new insights on these enigmatic sources. We show that the
nearly 100\% linear polarization and small variation of the
polarization position angles (PAs) of multiple bursts from the same source
suggest that the radiation is produced near the surface of a strongly
magnetized neutron star. 
As the emitted radiation travels through the magnetosphere, the
electric vector of the X-mode wave adiabatically rotates and stays
perpendicular to the local magnetic field direction. The PA freezes at
a radius where the plasma density becomes too small to be able to turn
the electric vector. At the freeze-out radius, the electric field is
perpendicular to the magnetic dipole moment of the neutron star
projected in the plane of the sky, independent of the radiation
mechanism or the orientation of the magnetic field in the emission
region. We discuss a number of predictions of the model. The variation
of PAs from repeating FRBs should follow the rotational period of the
underlying neutron star (but the burst occurrence may not be
  periodic). Measuring this period will provide crucial 
support for the neutron star nature of the progenitors of FRBs.  For
FRB 121102, the small range of PA variation means that the magnetic
inclination angle is less than about $20^{\rm o}$ and that the observer's line
of sight is outside the magnetic inclination cone. Other
repeating FRBs may have a different range of PA variation from that of
FRB 121102, depending on the magnetic inclination and the observer's
viewing angle.

\end{abstract}


\section{Introduction}

Fast radio bursts (FRBs) are millisecond-duration, bright, 
transient events, first detected at $\sim$GHz frequencies by the
Parkes Telescope during pulsar surveys
\citep{2007Sci...318..777L, 2013Sci...341...53T}. These bursts are
distributed roughly isotropically in the sky, and have an all-sky
rate of $\sim$$10^3$ to $10^4\rm\ day^{-1}$ above  
$\sim$$1\rm\ Jy\,ms$ fluence
\citep{2013Sci...341...53T, 2015MNRAS.447.2852K, 2016MNRAS.455.2207R,
  2016MNRAS.tmpL..49C, 2018MNRAS.475.1427B}. One object (FRB 121102)
has produced numerous  
bursts \citep{2016Natur.531..202S, 2016ApJ...833..177S}, and its
sky location has been determined with an accuracy of  
$\sim$$0.1^\pp$ by interferometry with the Jansky Very Large  
Array \citep{2017arXiv170101098C}.  The source is found to be 
associated with a dwarf star-forming host galaxy at redshift z =
0.19273 \citep{2017ApJ...834L...7T}. The FRB is also associated with a 
persistent radio source \citep{2017arXiv170101098C}, and their
projected separation is further pinned down to
$\lesssim0.01^\pp$ ($\lesssim 40\rm\ pc$ in physical distance) by the
European VLBI Network 
\citep{2017ApJ...834L...8M}.

The dispersion measure for FRBs, i.e. the number of free electrons per unit
area between the source and the Earth, is between $\sim$200 and
$2.5\times10^3\,$cm$^{-3}$pc\footnote{A list of all reported FRBs and their
  properties can be found at the FRB catalog
  \href{http://frbcat.org}{http://frbcat.org} \citep{2016PASA...33...45P}.}. 
For most of the reported bursts, this column density is larger than the
contribution from the  interstellar medium in the Milky Way by roughly
an order of magnitude, which suggests that these events lie at a
distance of a billion light years or more \citep[see][for recent
reviews]{2016MPLA...31..30013K, 2018arXiv180409092K}. This is confirmed
by the host galaxy 
of the repeater FRB 121102, which is at $\sim$$1\,$Gpc. If the source
radiates isotropically, then the energy release in the radio band for
bursts from the repeater FRB 121102 varies from $\sim$$10^{37}$ to
$10^{40}\,$erg \citep{2016Natur.531..202S, 2016ApJ...833..177S,
  2017ApJ...850...76L, 2017ATel10693....1O}. For other FRBs with
unknown distances, if the dispersion measure is dominated by the
intergalactic medium, then the isotropic equivalent energy release in the
radio band ranges from $\sim$$10^{39}$ to $10^{42}\,$erg (data from
the FRB catalog).

Polarization properties of many of the (so-far) non-repeating FRBs
have been measured \citep[e.g.][]{2015MNRAS.447..246P,
  2015Natur.528..523M, 2016Sci...354.1249R, 2017MNRAS.469.4465P,
  2018arXiv180409178C}, and the reported degree of linear polarization
is between $\sim$0\% and $\sim$80\%. However, the frequency
  resolution of most surveys is limited (e.g., channel width of 0.4
  MHz at 1.4 GHz), so the measurements could be affected by Faraday
  depolarization. Therefore, the above measurements of the degree of
  linear polarization should be viewed as lower limits.  Polarization
  of the repeater (FRB 121102) was measured at 4-8 GHz with high
accuracy for 26 bursts spread over 7 months in the rest frame of
the host galaxy \citep{2018Natur.553..182M, 2018arXiv180404101G}.  The
degree of linear polarization for all these bursts was nearly 100\%. Over
the duration of each burst, the polarization position angle (PA), or the
orientation of the electric field plane at frequency
$\nu$$\rightarrow$$\infty$, stayed nearly 
constant. However, the PA was found to vary from burst to 
burst but within a small range. The (error-weighted) 
standard deviation for the PAs of the 26 bursts detected is $\sigma \approx
11.5^{\rm o}$, so we have a crude estimate of PA 
variation range\footnote{The model presented in this paper does 
not depend on the precise value of $\Delta {\rm PA}_{\rm max}$, which
could be different for other repeating FRBs.} to be $\pm\Delta {\rm
PA}_{\rm max}\sim \pm2\sigma\sim \pm20^{\rm o}$. Moreover, the
rotation measure (RM) was found to be about 
$10^5$ rad m$^{-2}$, which varied from one burst to another by about
10\%.  The contribution to the RM from the Milky Way galaxy and the
intergalactic medium is estimated to be less than about 10$^2$ rad
m$^{-2}$ \citep{2018Natur.553..182M}, so most of the RM is likely to
be from the host galaxy and possibly the close vicinity of the burst;
the magnetic field strength inferred from the RM is, therefore, of
order milli-Gauss. The high degree of
linear polarization and the small range of PA variations among different
bursts over 7 months provide important constraints on the radiation
mechanism and the FRB source properties, which we address in this
paper.

\section{Polarization angle change as waves move through a neutron star magnetosphere} 

The sub-millisecond variability of FRB lightcurves
\citep[e.g. $\sim$30$\,\mu$s for
FRB170827,][]{2018arXiv180305697F} suggests that the 
underlying object is likely to be
a neutron star or a stellar-mass black hole\footnote{The
progenitors of FRBs could be much larger objects if bursts are produced in
small patches, or if the emission is produced by a source that is moving
toward the Earth at a speed close to the speed of light
\citep[e.g.,][]{narayan_kumar_09,lazar_etal_09,narayan_piran_12}.}. 
Considering that the brightness temperature of FRB radiation is
$\gtrsim$10$^{35}\,$K \citep{2014PhRvD..89..3009K,
  2014ApJ...785L..26L}, the radiation mechanism must be a coherent
process where particles in the 
source radiate in phase. Broadly speaking, this means that the radiation
mechanism is either a maser-type process (such as synchrotron, curvature,
cyclotron-Cherenkov, or other plasma masers) or the antenna
mechanism. Most maser and plasma processes require the wave
frequency to be closely  
related to either the plasma frequency or the cyclotron frequency, and
thus the source region should have a relatively low electron density 
(of order $10^{10}\gamma^{1/2}\,$cm$^{-3}$) and/or modest
magnetic field strength of order 10$^3\gamma\,$G
\citep{2018MNRAS.477.2470L}, where $\gamma$ is the typical Lorentz
factor of the radiating particles. These
conditions arise naturally far from the neutron star (or black
hole), near or beyond the light cylinder radius, $R_{\rm
  L} = c/\Omega$, where $\Omega$ is the rotational angular frequency
of the neutron star, and $c$ is the speed of light. The antenna 
mechanism, on the other hand, requires very strong 
magnetic field ($B\gae 10^{14}\,$G) and high electron density ($n\gae
10^{17}\,$cm$^{-3}$), which means that it can only operate within a
distance of a few $R_{\rm ns}$ from the surface of the neutron star
\citep{2017MNRAS.468.2726K}, where $R_{\rm ns}$ is the neutron star radius.

The small PA variation between bursts from the repeater
may require fine tuning for any mechanism that operates at large
distances from the neutron star or black hole,
because the magnetic field direction can change randomly in time and
space (and hence from burst to burst). In this paper, we assume that
FRBs are produced in the 
magnetosphere of neutron stars somewhere between the surface and the light
cylinder, where the magnetic field configuration is likely to be stable
(in the corotating frame of the neutron star) over
a long period $\gtrsim$7 months. For models in which the
bursts are generated near the surface, if the emitting region moves
around randomly between bursts, and hence samples different 
local magnetic field configurations (which would generally be a
superposition of dipole plus higher order multipoles and could,
in principle, be arbitrarily complicated), it might seem that the PA
should correspondingly change by a large amount between bursts.
This turns out not to be the case. 

We show that the PA of bursts generated near the neutron star
surface is modified as the wave travels through the magnetosphere in such a way that
the escaping radiation is nearly 100\% linearly polarized,
with the electric vector always perpendicular to the
magnetic dipole moment projected in the plane of the sky. Thus, the
PAs of different bursts are not expected to vary by much more than the angle
between the rotation axis and magnetic axis of the neutron star (provided the
observer is outside the magnetic inclination cone, see \S 3). This result
does not depend on the radiation mechanism or the magnetic field
direction at the source position. 

In the following, we first describe the basic picture of radio waves
propagating through a neutron star magnetosphere (\S 2.1) and then
solve the linear wave equation for the change of PA along the
propagation path (\S 2.2). Then in \S 2.3, we extend the discussion to
the high intensity FRB waves in the non-linear regime
and calculate the location beyond which the magnetosphere cannot
supply sufficient current density and hence the PA freezes.

\subsection{The Basic Picture}

Consider an electromagnetic (EM) wave of frequency $\omega$
propagating in a quasi-static medium with magnetic field $B$ and
plasma frequency $\omega_{p}$; $\omega_{p}^2 = 4\pi q^2 \sum_s
(n_s/m_s)$, where $q$, $m_s$ and $n_s$ are the charge, mass and number
density of electrons ($s = -$) and protons/positrons ($s = +$). The
magnetosphere of a neutron star is usually not charge neutral
everywhere, and hence $n_+\neq n_-$.  It is convenient to express the
two polarization states of EM radiation in the presence of a strong
magnetic field (to be quantified shortly) as those with electric
vector perpendicular to the plane of the magnetic and wave vectors (called
the X-mode) and those with electric vector in this plane (the O-mode). The
cyclotron frequency of electrons (or positrons) 
is $\omega_B = qB/(m c)$ and the medium of interest to us has a strong
magnetic field with $\omega_B \gg \omega$. It can be shown
  \citep{1986ApJ...302..120A} that, if $\omega_{p}\gg \omega$ near
  the source region, the O-mode waves\footnote{Throughout this paper,
    the waves with electric vector in the $\vec{k}$--$\vec{B}$ plane
    are in general referred to as the O-mode regardless of the order
    of $\omega$ and $\omega_p$. In the literature, the wave branch
    with $\omega\ll \omega_p$ is sometimes also called the
    Alfv{\'e}n-mode.}  propagate along the local magnetic field and
either hit the neutron star surface (along closed field lines) or
are strongly diluted as they propagate outwards due to rapidly
diverging magnetic field (wave energy flux dropping $\propto r^{-3}$
or faster). Based on this consideration, we assume that a large fraction
of the source emission is initially in the form of the X-mode and we
focus on the propagation  of X-mode waves through the magnetosphere of
the neutron star. Our general consideration applies to, but is not
  limited to, the coherent curvature emission mechanism, which
  generates primarily X-mode waves \citep{2017MNRAS.468.2726K}.

\begin{figure}
  \centering
\includegraphics[width = 0.48 \textwidth,
  height=0.25\textheight]{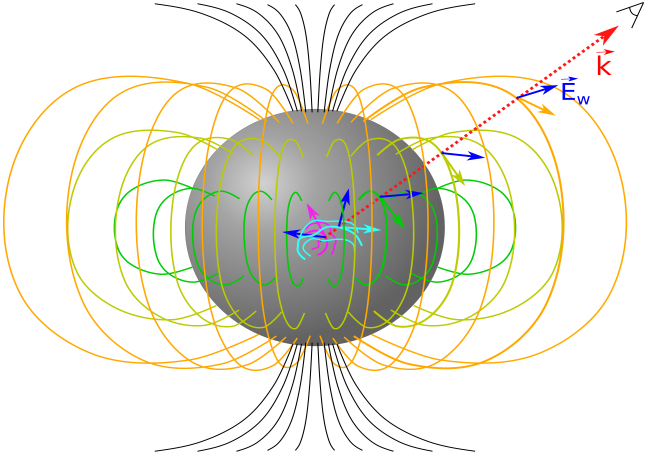}
\caption{Physical picture of an X-mode EM wave traveling through
  a neutron star magnetosphere. The magnetic field configuration
  near the source region, which is near the surface, is complex. As
  the wave travels outwards, the electric vector adiabatically rotates in such a way
  that it is always perpendicular to the local magnetic field. At
  large distances from the neutron star (but still well below the
  light cylinder), where the plasma density
  becomes too small to force the electric vector to follow the
  changing direction of the magnetic field, the polarization position angle is
  frozen. At this freeze-out radius (not shown in this picture since
  it is $\gg R_{\rm ns}$), the magnetic field is nearly
  dipolar and the wave vector is nearly in the radial direction from
  the center of the star, so the electric vector of the escaping wave
  is perpendicular to the magnetic dipole moment of the neutron
  star.
}\label{fig:frb-fig1}
\end{figure}

We show that, under suitable conditions, as the wave travels to larger 
distances from the source, its electric vector adiabatically rotates in 
such a way that it stays perpendicular to the local
magnetic field (and also the wave vector) as shown in 
Fig. \ref{fig:frb-fig1}.
\citet{1979ApJ...229..348C} considered this ``adiabatic
walking" phenomenon in the context of radio emission from regular
pulsars. In this paper, we show that the same phenomenon is
  applicable for much stronger magnetic fields (e.g. near a magnetar) and
  for radio waves with much higher intensity as inferred from FRBs.
Beyond a certain distance from the neutron star surface the PA 
of the wave is frozen; this occurs when the plasma density is too
small to be able to force the electric vector to follow the turning
and twisting magnetic field lines.

The physics behind adiabatic walking of an EM wave traveling through a
neutron star magnetosphere is as follows. The EM wave induces 
current oscillations, which in turn generate
magnetic and electric fields oscillating at the 
same frequency as the wave. The induced electric field is nearly
parallel to the local, non-oscillatory, magnetic field and its 
magnitude is such that, when appropriate conditions are satisfied
as described below, the superposition of the two
electric fields remains perpendicular to the local magnetic field.

\subsection{Wave Propagation through the Magnetosphere --- Linear
  Regime}

In this subsection, we consider the propagation of a weak EM wave
in the linear regime where particles in the magnetospheric
plasma have non-relativistic speeds. Since in this regime the wave
equation can be easily solved analytically and numerically,
interesting insight can be gained on the physics of the adiabatic
walking. Additional discussion of the linear regime can be found in Appendix~A. We extend the discussion to the non-linear regime in 
the next subsection.

In the following, we decompose vectors into components parallel
and perpendicular to the local static magnetic field $\vec{B}$,
and identify these components by subscripts $\parallel$ and $\perp$, respectively.
The velocity of a particle of charge $q$ and mass $m$ in the presence of a 
time-varying wave electric field $\vec{E}_{w}\mr{exp}(-i\omega t)$ and
a static magnetic field $\vec{B}$ is $\vec{v}\mr{exp}(-i\omega t)$
(the Fourier components at other frequencies are not relevant), where
the components of $\vec{v}$ are given by
\begin{equation}
\label{eq:particle-speed}
  v_\parallel = {iq E_{w\parallel}\over m\omega}, \
  \vec{v}_\perp = {q \left[ -i\omega
     \vec{E}_{w\perp} + w_B \vec{E}_w\times\hat{B}\right] \over
   m(\omega_B^2 - \omega^2)}.
\end{equation}

For a rotating force-free magnetosphere with angular velocity 
$\vec\Omega$, the Goldreich-Julian
number density is equal to \citep{1969ApJ...157..869G}
\begin{equation}
n_{\rm GJ} = \frac{\vec{\Omega}\cdot \vec{B}}{2\pi qc}.
\end{equation}
However, the actual number density of the plasma, $n =                  
n_+ + n_-$, will generally be larger than
this by a multiplicity factor $\mc{M}$, i.e.
\begin{equation}
n = \mc{M} n_{\rm GJ}.
\end{equation}
Equivalently, for an electron-position
plasma with total number density $n$, the net charge density is equal
to $nq/\mc{M}$.  For the case of a neutron star magnetosphere, where $\omega\ll
\omega_B$, we can ignore the  
second-order small terms $\mc{O}(\omega^2/\omega_B^2)$ in
eq. (\ref{eq:particle-speed}).
The components of the current density vector
$\vec{j}$ can then be obtained from the velocity components
in eq. (\ref{eq:particle-speed}):
\begin{equation}
\label{eq:2}
   j_\parallel = {i\omega_p^2 E_{w\parallel}\over 4\pi\omega}, \
   \vec{j}_\perp \approx -{i\omega\omega_p^2\over 4\pi \omega_B^2} 
       \left(\vec{E}_{w\perp} + {i\omega_B\over \mc{M}\omega} \vec{E}_{w}
       \times \hat{B}\right),
\end{equation}
or
\begin{equation}
  \label{eq:current}
  {4\pi i\over \omega}\vec{j} = -{\omega_p^2\over \omega^2}E_{w\parallel}\hat
    B + {\omega_p^2\over \omega_B^2} \vec{E}_{w\perp}+ {i\omega_p^2 \over \mc{M} 
     \omega\omega_B} \vec{E}_{w} \times \hat{B}.
\end{equation}
This current affects the wave electric field 
along the propagation path according to the following 
wave equation (which is obtained by combining
Faraday's and Ampere's laws)
\begin{equation}
    {c^2\over \omega^2}\nabla\times(\nabla\times\vec{E}_w) - \vec{E}_w= {4\pi
    i \over \omega}\vec{j}.
  \label{eq:waveEq1}
\end{equation}

In the following, we first show that the second and third current terms in
eq. (\ref{eq:current}) are negligible. The ratio $\omega_p^2/\omega_B^2$ is
equal to the reciprocal of the plasma magnetization parameter 
($\sigma \equiv B^2/4\pi nmc^2$). In the neutron star magnetosphere, we find
\begin{equation}
  \label{eq:24}
  \omega_p^2/\omega_{\rm B}^2\sim
  10^{-15}\mc{M}B_{0,15}^{-1}P^{-1}(r/100R_{\rm ns})^3 \lll 1,
\end{equation}
where
$B_0$ is the surface dipole magnetic field (we assume
$B\propto r^{-3}$ for $r\gg R_{\rm ns}$). Throughout this paper, we use
$B_0 = 10^{15}\,$G as our fiducial value, motivated by the inferred
magnetic field strength of nearby magnetars
\citep[e.g.][]{2017ARA&A..55..261K}. By equation (\ref{eq:24}) we see that the second
term in the current equation (\ref{eq:current}) is always
negligible. The third term ($\propto \vec{E}_w\times\hat{B}$) gives
rise to Faraday rotation. Since left and right circularly polarized
waves propagate at different phase speeds, the polarization plane rotates by $\Delta
\phi = R\omega_p^2 /(\mc{M}\omega_Bc) = 4\pi R n q/\mc{M}B$ after a
propagation length of $R$. Since $n/\mc{M} = n_{\rm GJ} \sim
B\Omega/(2\pi qc)$, we find $\Delta \phi \sim 
2R\Omega/c = 2R/R_{\rm L}$. In \S2.3 and \S3, we show that the small
range of PA variations for FRB 121102 requires the exit
point (where the adiabatic walking behavior stops) to be much below
the light cylinder radius $R_{\rm L}$. Hence, $\Delta \phi\ll 1\,$rad, and the
third term in equation (\ref{eq:current}) can also be
ignored. We include this term in Appendix B for our numerical
solutions to the full wave equation and find that its effect on the PA
change along the path of the wave is negligible. 
We are thus left with only the first term in equation (\ref{eq:current}).

An X-mode EM wave has, by definition, electric field
$\vec{E}_{w}$ perpendicular to the $\vec{k}${--}$\vec{B}$ plane,
  hence $E_{w\parallel}=0$. If the wave propagates in a
  \textit{homogeneous} plasma, where the magnetic field is uniform,
  the electric vector remains perpendicular to the
  $\vec{k}${--}$\vec{B}$ plane as the wave propagates, and the first
  term on the right hand side of equation (\ref{eq:current}) also
  vanishes. Thus, the entire right side of equation (\ref{eq:current})
  is essentially zero, and the wave propagates in a straight line with
  refractive index $\approx 1$, as if in vacuum (all source terms are
  negligible). In particular, the PA remains constant.

\begin{figure*}
  \centering
\includegraphics[width = 0.53 \textwidth,
 height=0.28\textheight]{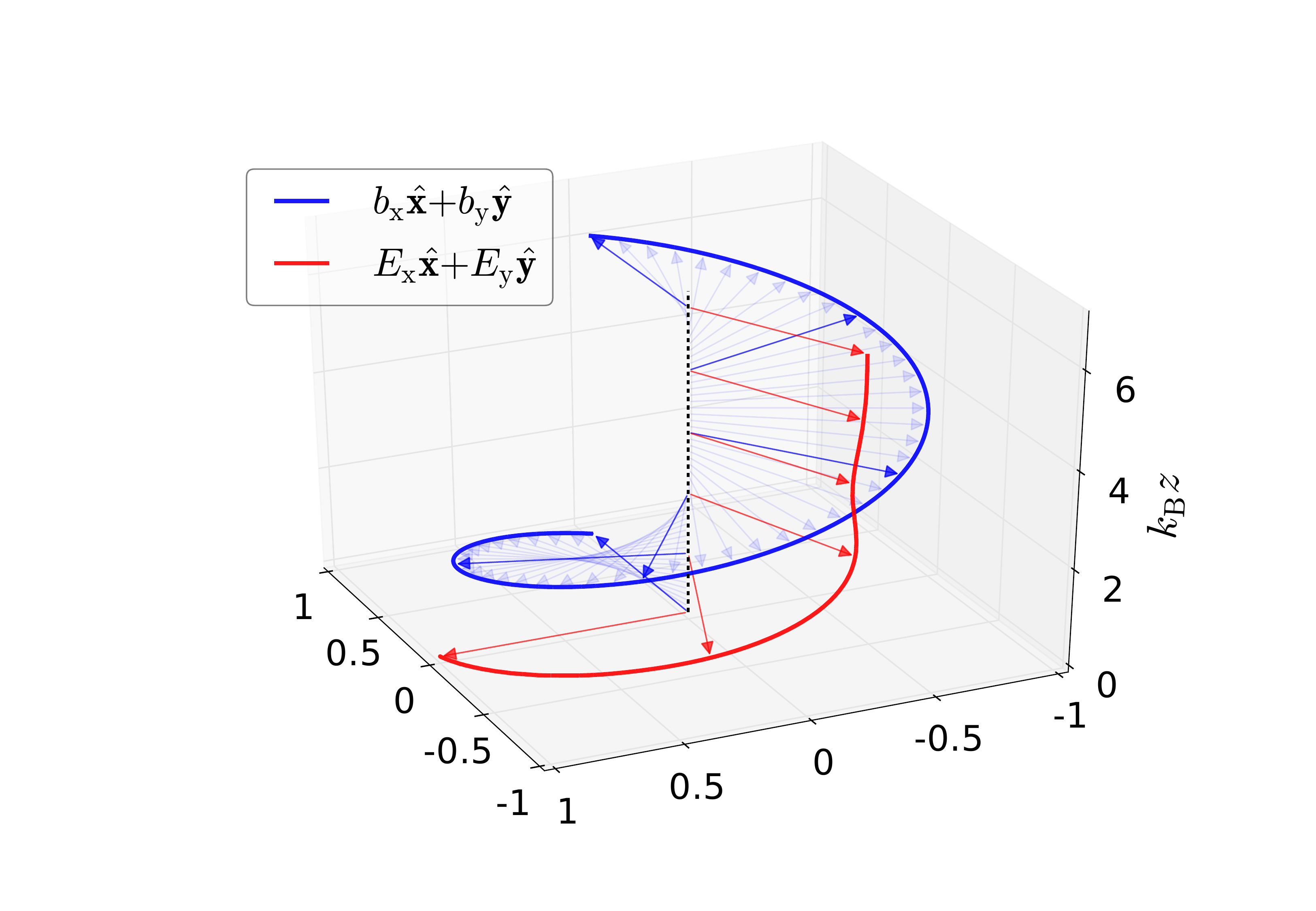}
\includegraphics[width = 0.42 \textwidth,
 height=0.23\textheight]{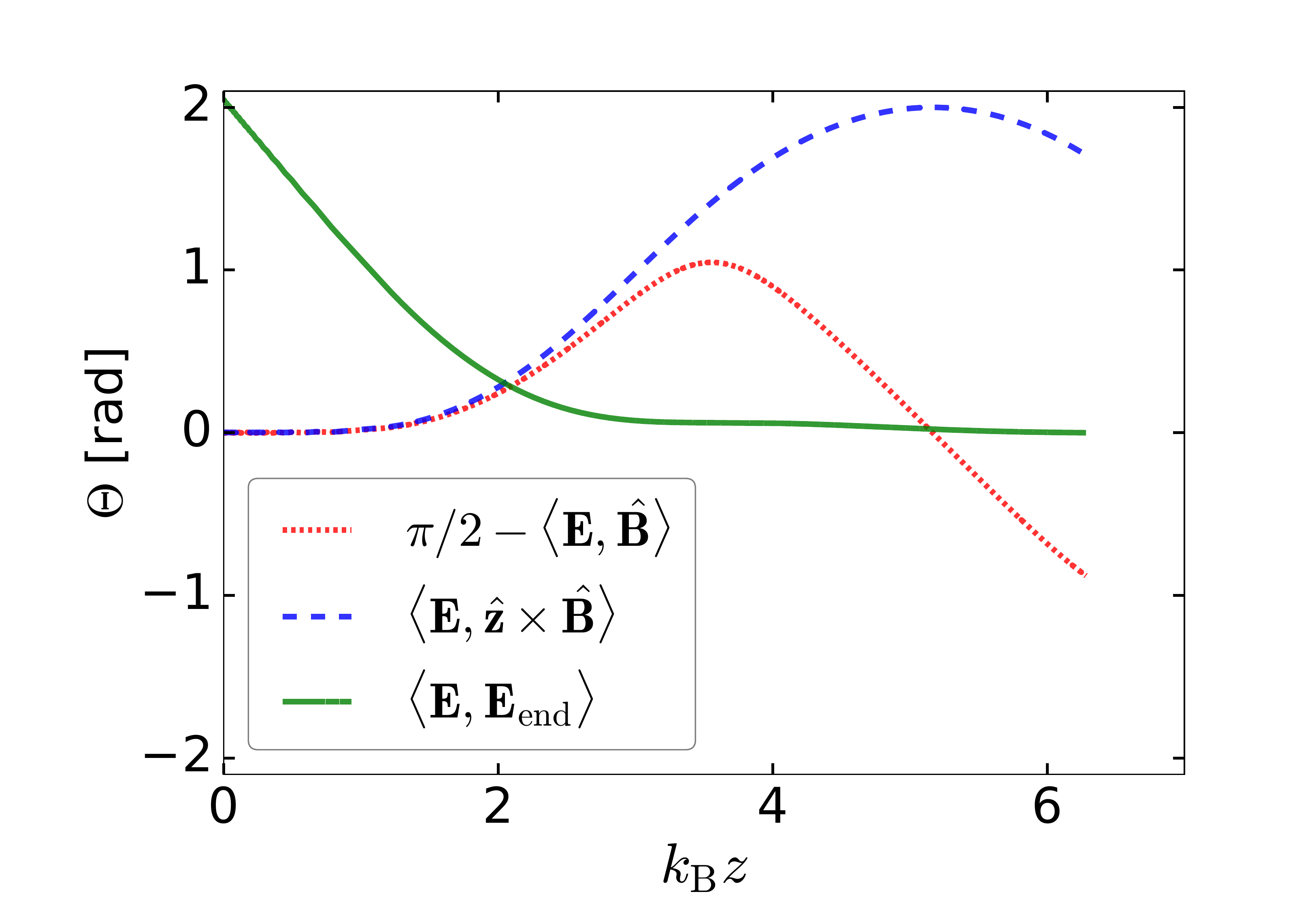}
\caption{Numerical integration of the 1-D linear wave equation
  (\ref{eq:100}) for an example in which the magnetic field rotates
  around the z-axis at a rate $k_{\rm B}\equiv R_{\rm B}^{-1}$ with
  pitch angle $\theta = 60^{\rm o}$:
 $\hat{B} = \sin\theta \cos k_B z\, \hat{x} +
\sin\theta \sin k_Bz\, \hat{y} + \cos\theta \,\hat{z}$.
 The boundary conditions are: $\vec{E}_{\rm w}(z=0) = \hat{y}$
  (normalized so that $|E_{\rm w}|=1$) and $\d \vec{E}_{\rm w}(z=0)/\d
  z_1 = 0$. The plasma density is taken to vary with $z$ as:
  $\omega_p^2k/\omega^2 k_{\rm B}= 1000(k_{\rm B}z + 1)^{-5}$, where
  $k\equiv \omega/c$. \textit{Left
      Panel:} The magnetic field projected in the x--y plane is shown
    as blue arrows, and the wave electric field is shown as red
    arrows. \textit{Right Panel:} The angle
    between the wave electric vector and the static magnetic field is
    shown as red dotted line, the angle between the wave electric
    vector and $\hat{z}\times \hat{B}$ is shown as blue dashed line,
    and the angle between the wave electric vectors at current 
    location $\vec{E}(z)$ and at the right-hand boundary 
    $\vec{E}_{\rm end}(k_{\rm B}z_{\rm end}= 2\pi)$ is shown as
    green solid line. The angle between two vectors $\vec{A}_1$ and  
    $\vec{A}_2$ is in general denoted as $\langle \vec{A}_1, \vec{A}_2
    \rangle\equiv \mr{arccos}(\vec{A}_1\cdot \vec{A}_2/A_1A_2)$. When
  $\omega_p^2k/\omega^2k_{\rm B}\gg 1$ ($k_{\rm 
      B}z\ll 1$), the wave electric vector adiabatically rotates along
    with the magnetic field, keeping $\vec{E}_w$ nearly aligned with
    $\vec{k}\times\vec{B}$. As the plasma density drops, when
    $\omega_p^2k/\omega^2k_{\rm B}\lesssim 2$ ($k_{\rm B}z\gtrsim 2$),
    the wave electric vector stops rotating and the PA freezes
    (i.e. the wave propagates as if in vacuum).
}\label{fig:var}
\end{figure*}

The situation is different if we consider an inhomogeneous
plasma, in which the direction of the magnetic field changes (by
$\sim$1$\,$rad) on a length-scale $R_{\rm B}\gg \lambda$, where
$\lambda$ is the wavelength of the EM wave. The plasma density may
also change on a similar length-scale. Now, even if a wave starts off
with $\vec{E}_{w}$ perpendicular to the  $\vec{k}${--}$\vec{B}$ plane,
it will not remain so, because the magnetic field $\vec{B}$ rotates
from its initial direction as the wave propagates. As a result, there
will be a non-zero value for the first current term in
eq. (\ref{eq:current}), which is parallel to the local magnetic
field. This current induces an electric field which is also parallel
to the local magnetic field. The superposition of the induced field
and the original wave field  causes the electric vector to rotate.
As we show, the rotation is such that the wave polarization
follows the twisting magnetic field as the wave propagates through the
magnetosphere.

For simplicity, we consider a plane wave with electric
field amplitude in the form $\vec{E}_w(z_1)\mr{e}^{iz_1}$, where $z_1
= kz$ and $k \equiv \omega/c$ is a constant. Then the wave equation 
(\ref{eq:waveEq1}) becomes
\begin{equation}
  \label{eq:6}
  \begin{split}
    2i {\d \over \d z_1}(E_{\rm x}\hat{x} +
E_{\rm y}\hat{y}) &= -{4\pi i \over
  \omega}(j_{\rm x}\hat{x} + j_{\rm y} \hat{y}) \mr{e}^{-iz_1},\\
 E_{\rm z}\hat{z} &= -{4\pi i\over
  \omega} j_{\rm z} \mr{e}^{-iz_1},
  \end{split}
\end{equation}
where we have ignored the second-order derivative terms $\d^2/\d
z_1^2\sim \mc{O} (k^2R_{\rm B}^2)$ (they are included in the
numerical calculations in Appendix B). We further simplify the
calculation by assuming that the magnetic field is in the x{--}y plane
$\hat{B}(z_1) = \cos k_B z\, \hat{x} + \sin k_Bz\, \hat{y}$ (where
$k_B\equiv R_{\rm B}^{-1}$) and hence $E_{\rm
  z}\equiv 0$ (as we show in Appendix B, the qualitative
  behavior of adiabatic walking is the same for an arbitrary magnetic
  field orientation). Then we make use of the current density 
$4\pi i\omega \vec{j}\approx -\omega_p^2E_{w\parallel}\hat{B}
\mr{e}^{iz_1}$ (from eq. \ref{eq:current}) and the wave equation
becomes 
  \begin{equation}
    \label{eq:15}
    2 i{\d \vec{E}_{\rm w} \over \d z_1} \approx {\omega_{\rm p}^2 \over
      \omega^2} (\vec{E}_{\rm w}\cdot \hat{B})\hat{B}.
  \end{equation}
Multiplying (dot-product) eq. (\ref{eq:15}) by 
$\hat{B}$, we obtain
\begin{equation}
  \label{eq:9}
 {\d \over \d z_1} (\vec{E}_w\cdot \hat{B})  \approx {\omega_p^2\over
   2i\omega^2} (\vec{E}_w\cdot \hat{B}) + \vec{E}_w\cdot {\d
   \hat{B}\over \d z_1}.
\end{equation}
Since the derivative term on the right hand side can be roughly written as
$\vec{E}_w\cdot (\d \hat{B}/\d z_1)\sim E_w/kR_{\rm B}$, the 
equation above can be approximately integrated analytically. Using
the X-mode boundary condition $\hat{E}_w\cdot \hat{B}|_{z=0} = 0$,
we obtain
\begin{equation}
  \label{eq:10}
\mr{Re}\left( \hat{E}_w\cdot \hat{B}\right) \sim 
{2\omega^2 \over kR_{\rm B}\omega_p^2} \sin {kz\omega_p^2 \over
  2\omega^2}.
\end{equation}
We see that the angle between the wave electric vector and the
magnetic field oscillates with a spatial wavelength of $2\pi\omega
  c/\omega_p^2$ and an amplitude $\sim$$\omega^2/(kR_{\rm
    B}\omega_p^2)$. The electric field will follow the direction of the
magnetic field so long as the amplitude of the above oscillation is
less than about a radian. 
Therefore, the direction of the electric vector rotates and stays
approximately perpendicular to the local magnetic field as long
as
\begin{equation}
\label{eq:freezeout1}
   {\omega_p^2\over \omega^2} \gtrsim {\lambdabar\over R_{\rm B}},\
   \ \lambdabar = \frac{1}{k}= {\lambda\over 2\pi}. 
\end{equation}

A more precise analytical solution of the wave equation under
arbitrary boundary conditions is provided in Appendix A, where we
show, using an eigenmode analysis, how the wave electric vector rotates as
the magnetic field turns. In Appendix B, 
we present results from numerical integration of the wave equation
when $\omega_p$ varies with $z$. One example of such a 
numerical integration is shown in Fig. \ref{fig:var}. 
Here, the magnetic field (only the x--y components are shown as blue arrows)
rotates around the z-axis at a rate $k_{\rm B}\equiv R_{\rm B}^{-1}$ with pitch angle
$\theta = 60^{\rm o}$: $\hat{B} = \sin\theta \cos k_B z\, \hat{x} +
\sin\theta \sin k_Bz\, \hat{y} + \cos\theta \,\hat{z}$. The boundary
condition is that the wave 
starts at $z=0$ as a pure X-mode $\vec{E}_w(z=0)=\hat{y}$. The plasma
density decreases with z as $\omega_p^2k/\omega^2 k_{\rm B}=  
1000(k_{\rm B}z + 1)^{-5}$, with $\omega_p^2$ and $k_{\rm
  B}$ dropping as $r^{-4}$ and $r^{-1}$ (as a simple example). In a realistic
neutron star magnetosphere, the scaling of
$\omega_p^2{k}/\omega^2k_{\rm B}$ as a function of radius may be more
complicated, but the qualitative result should be the same. For the example shown, we
see that the electric vector (red arrows) stays perpendicular to the
rotating magnetic field until $\omega_p^2/\omega^2 \simeq
2\lambdabar/R_{\rm B}$ (or $k_{\rm B}z\simeq 2$), beyond which the
wave approaches vacuum-like propagation.


In a neutron star magnetosphere, the condition (\ref{eq:freezeout1})
for adiabatic walking of $\vec{E}_w$ is violated at some distance from
the surface, the freeze-out radius $\t{R}_{\rm fo}$, beyond which the
change of PA is negligible. We assume that the magnetic field far from the
neutron star surface, but well within the light cylinder ($R_{\rm L}=c/\Omega$), is
dominated by the dipole component $B(r)\propto r^{-3}$, as higher order
multipoles decline more rapidly. The particle density is a multiple $\mc{M}$
of the Goldreich-Julian density $n_{\rm GJ}$, so the
freeze-out radius is
\begin{equation}
\label{eq:Rfo1}
   {\t{R}_{\rm fo}\over R_{\rm ns}} \simeq 3.4\times10^4 \left[{ B_{0, 15}
      \mc{M} (R_{\rm B}/\t{R}_{\rm fo})
       \over \nu_9P}\right]^{1/2},
\end{equation}
\begin{equation}
  \label{eq:28}
   {\t{R}_{\rm fo}\over R_{\rm L}} \simeq 7 \left[{ B_{0, 15}
      \mc{M} (R_{\rm B}/\t{R}_{\rm fo})
       \over \nu_9P^3}\right]^{1/2}.
\end{equation}

We see that this estimate of the freeze-out radius in the linear
regime may be close to or beyond the light cylinder. At these radii,
the magnetic field is dominated by the 
toroidal rather than the dipolar component of the field, so the PA
may vary by a large amount from one burst to
another. However, as we show in the next subsection, the adiabatic
walking condition for large amplitude FRB waves --- in the non-linear
regime --- is quite different.

\subsection{Wave Propagation through the Magnetosphere --- Non-linear
  Regime}

A more stringent condition than eq. (\ref{eq:freezeout1}) needs to be
satisfied in the case of the high intensity 
waves present in FRBs. For waves with $q E_{w\parallel}/
 m c\omega > 1$, the particle velocity component
parallel to the local magnetic field becomes relativistic. When
$E_{w\perp}\gtrsim B$, the transverse velocities also become
relativistic (see eq. \ref{eq:particle-speed}). Relativistic motions
break the linear relation between current density and the wave
amplitude, so the wave equation becomes highly complicated
(numerical simulations of particles' orbits are needed to calculate
the correct current density). In this section, we provide a rough
analytical estimate of the condition at which adiabatic walking
is terminated for large amplitude EM waves.

Large amplitude EM waves require a large current in order to rotate
$\vec{E}_w$ and keep it perpendicular to the changing
$\vec{k}${--}$\vec{B}$ plane. However, the maximum current density
(parallel or perpendicular to the magnetic field)
that can be supplied by the plasma is $j_{\rm max}\simeq
qnc$. If this maximum current
oscillates at frequency $\omega$, then the maximum amplitude of the
current term on the right-hand side of the wave equation (\ref{eq:6}) is
\begin{equation}
  \label{eq:8}
  {4\pi\over \omega}j_{\rm max} \simeq {\omega_{\rm p}^2 \over
    \omega^2} {mc\omega\over q} = {\omega_{\rm p}^2\over
    \omega^2}{E_{\rm w}\over a_0},
\end{equation}
where $a_0 = E_{\rm w}q/mc\omega$ is known as the non-linearity
parameter of the EM wave. The wave amplitude at radius $r$ ($\gg$ the
source size) from the source is $E_{\rm w} = \sqrt{L/r^2 c}$, where $L$ is
the isotropic equivalent luminosity of an FRB. Hence,
\begin{equation}
\label{eq:3000}
a_0 \equiv {qE_{\rm w}\over mc\omega} \simeq 5.1\times10^5
\frac{L_{41}^{1/2}}{r_7 \nu_9} \gg 1, 
\end{equation}
where we use the typical isotropic equivalent luminosity $L =
10^{41}L_{41}\rm\,erg\,s^{-1}$ of bursts detected by
\citet{2018Natur.553..182M, 2018arXiv180404101G} as a fiducial
value. 

We take the absolute value of xy
components of the wave equation (\ref{eq:6}) and make use of the
maximum current in eq. (\ref{eq:8}), and obtain
\begin{equation}
  \label{eq:5}
  2\left|{\d \over \d z_1}(E_{\rm x}\hat{x} +
E_{\rm y}\hat{y}) \right| = {4\pi \over \omega} |j_{\rm x}\hat{x} +
j_{\rm y}\hat{y}|\lesssim {\omega_{\rm p}^2\over \omega^2}{E_{\rm
    w}\over a_0}.
\end{equation}
To keep $\vec{E}_w$ perpendicular to (for X-mode) or inside (for
O-mode) the rotating $\vec{k}$--$\vec{B}$ plane, 
we require $\d (E_{\rm x}\hat{x} +E_{\rm y}\hat{y})/\d z_1\sim E_{\rm
  w}/kR_{\rm B}$. Thus, we obtain the condition for adiabatic walking
\begin{equation}
\label{eq:adiabatic-walking}
   {\omega_p^2\over\omega^2} \gtrsim {a_0 \lambdabar\over R_B},
\end{equation}
which is more stringent than the condition (\ref{eq:freezeout1})
since $a_0\gg 1$ (see eq. \ref{eq:3000}). Thus, the freeze-out radius
$R_{\rm fo}$ in the non-linear regime is
\begin{equation}
  \label{eq:Rfo2}
    {R_{\rm fo}\over R_{\rm ns}} \simeq 230 {B_{0,15} \mc{M} (R_{\rm
      B}/R_{\rm fo}) \over PL_{41}^{1/2}},
\end{equation}
\begin{equation}
  \label{eq:Rfo3}
  {R_{\rm fo}\over R_{\rm L}} \simeq 4.8\times10^{-2}\, {B_{0,15}
     \mc{M} (R_{\rm B}/R_{\rm fo}) \over P^2 L_{41}^{1/2}}.
\end{equation}
We see that $R_{\rm fo}$ here is much smaller than its counterpart in the
linear regime $\t{R}_{\rm fo}$ (eq. \ref{eq:Rfo1}). This is because the
extremely high intensity of FRB waves enables them to force their way through
the magnetosphere with no change in their PA when the induced electric
field by the plasma current is weaker than 
the parallel component of the wave electric field.


The length scale $R_{\rm B}$ over which the orientation of the magnetic field
changes by $\sim$1$\,$rad along the $\vec{k}$ direction depends on
the detailed magnetic field configuration in the magnetosphere. In the above
estimate, we adopted $R_{\rm B}=R_{\rm fo}$ as our fiducial
value but left $R_{\rm B}$ as a free parameter. Near the
neutron star surface ($r\sim \mbox{a few}\times R_{\rm ns}$) or near
the light cylinder 
($r\sim R_{\rm L}$), we may reasonably expect $R_{\rm
  B}\sim r$. However, this may not hold for $R_{\rm ns}\ll r\ll R_{\rm
  L}$. For instance, for a magnetosphere with only dipole and
quadruple components of equal strength near the surface, the dipole
component dominates at increasingly larger radius, and we expect
$R_{\rm B}/r\sim r/R_{\rm ns}$ until the toroidal component kicks in
near the light cylinder, where we have $R_{\rm B}\sim R_L$. Thus, for
this particular case, we have 
$R_{\rm B}\sim \mr{min}(r^2/R_{\rm ns}, R_{\rm L})$. However, without
a detailed model of the magnetosphere, the profile $R_{\rm B}(r)$ is
currently highly uncertain. Fortunately, as we show in \S3,
this does not lead to a large uncertainty on the PA of FRBs, as long
as $R_{\rm ns}\ll R_{\rm fo}\ll R_{\rm L}$.

We note that the freeze-out radius in eq. (\ref{eq:Rfo2}) is frequency
dependent, if the intrinsic emission spectrum $L(\nu)\sim \nu L_\nu$ is
not flat. Since $R_{\rm fo}\propto L(\nu)^{-1/4}$, the dependence of PA
on frequency is rather weak but may still be measurable in the
future. For the speculative case $R_{\rm B}\sim \mr{min}(r^2/R_{\rm ns}, R_{
\rm L})$ considered above and $\nu L_\nu\propto \nu^{\alpha}$, a rough
estimate of the PA variation is
\begin{equation}
  \label{eq:25}
  \nu {\d\mr{PA}\over \d \nu} \sim {\alpha \over 4} \mr{max}\left( {R_{\rm ns}\over
      R_{\rm fo}}, {R_{\rm fo}\over R_{\rm L}} \right),
\end{equation}
which gives $\Delta \mr{PA}\gtrsim (|\alpha|/4)\sqrt{R_{\rm ns}/R_{\rm
    L}} \sim 0.3^{\rm o} P^{-1/2}(\Delta \nu/\nu)$, for
$|\alpha|\sim\,$order unity. Since the profile $R_{\rm B}(r)$ is highly uncertain, 
the estimate above is approximate and should be taken with
caution. However, the qualitative prediction is that the wavelength
dependence of the measured plane of polarization should deviate
from the standard $\Theta(\nu)-\Theta(\infty)\propto \nu^{-2}$. A
simple way to test this is to measure the PAs with high accuracy at
two widely separated frequency bands (e.g. 4 and 8 GHz).

We note that the condition for adiabatic walking in
eq. (\ref{eq:adiabatic-walking}) and the freeze-out radius $R_{\rm
  fo}$ in eq. (\ref{eq:Rfo2}) apply to O-mode as well as X-mode
waves. As we show in \S 3, the small range of PA variation for FRB
121102 requires that the source location to be much below the
freeze-out radius $R_{\rm s}\ll R_{\rm fo}$ and the freeze-out radius
to be much below the light cylinder $R_{\rm 
  fo}\ll R_{\rm L}$. Near the source location (typically $\omega_p\gg
\omega$), the induced electric field associated with the 
plasma current (the first term in eq. \ref{eq:current}) cancels with
the parallel component of the wave electric field, and hence the wave
electric vector $\vec{E}_w$ for the O-mode is in the
$\vec{k}${--}$\vec{B}$ plane and stays 
perpendicular to $\vec{B}$. The wave magnetic field $\vec{B}_w$ is
perpendicular to the $\vec{k}${--}$\vec{B}$ plane, so we find that
the Poynting vector points in the direction of $\vec{E}_w\times
\vec{B}_w\para \vec{B}$, i.e. O-mode waves propagate along the local
magnetic field near the source region. This can also be seen by
calculating the group velocity from the O-mode dispersion relation
\citep[e.g.][]{1986ApJ...302..120A}. If the source generates
comparable amount of X- and O-mode waves, the high density plasma
near the neutron star surface (with $\omega_p$, $\omega_B\gg \omega$)
acts like a polarizing beam splitter which efficiently separates the
X-mode (propagating along a straight line) and the O-mode (propagating
along the local magnetic field line). If the wave amplitude is small
\citep[the limit considered by][in the radio pulsar
context]{1986ApJ...302..120A}, as the plasma density decreases towards
large radii, the O-mode phase velocity becomes sub-relativistic
(below the typical thermal speed of particles) and hence the wave is
severely Landau damped. However, since the extremely intense FRB waves
have energy density much larger than what
particles (with negligible inertia, see eq. \ref{eq:24}) can absorb,
the O-mode component may still be able to escape, as long as the magnetic
field line extends beyond the freeze-out radius in
eq. (\ref{eq:Rfo2}). Still, the energy flux of O-mode waves is
severely diluted as the magnetic field lines diverge towards
larger radii (the solid angle of the beam increases as $\propto r$ for a
pure dipole or faster if higher-order multipoles dominate), so the
observer may be biased towards detecting the undiluted, much brighter
X-mode flux.

\section{FRB Polarization, Progenitor Star Properties, and Radiation
  Mechanism} 
Let us consider the FRB source region to be located at a radius
$R_{\rm s}$. The physics of wave propagation in the neutron star
magnetosphere shown in \S2 suggests that the EM
waves reaching the observer correspond to the X-mode and will be
nearly 100\% linearly polarized. 
As long as the source is located well below the freeze-out radius
$R_{\rm s} \ll R_{\rm fo}$, and $\omega_{\rm p},\,\omega_{\rm 
  B}\gg\omega$ near the source, the
 PA measured by an observer at infinity has no memory of the magnetic
 field configuration at the source but 
reflects the direction of the magnetic field at $R_{\rm fo}$; the wave
electric vector is parallel to $\hat{k}\times\hat{B}$ at $R_{\rm
  fo}$. If $R_{\rm s} \ll R_{\rm fo}$ by a factor of 10 or more, then
the wave vector at the freeze-out radius is very nearly in the radial
direction. In this 
paper, we assume that the magnetic field at radius $R_{\rm ns} \ll r
\ll R_{\rm L} \equiv c/\Omega$ is nearly dipolar, $\vec{B}(\vec{r})
\simeq [3 \hat{r} (\hat{r}\cdot\vec{m}) - \vec{m}]/r^3$, where
$\vec{m}$ is the magnetic dipole moment of the neutron star. For
$R_{\rm ns} \ll R_{\rm fo} \ll R_{\rm L}$, the
direction of the wave electric field at $R_{\rm fo}$ is parallel to
$\hat{k}\times\hat{B} \parallel \hat{k}\times\hat{m}$, i.e. the PA is
perpendicular to the neutron star's magnetic moment projected in the
plane of the sky. For repeating bursts from the same object, the
PA would vary from one burst to another, if the magnetic and 
rotation axes are not parallel.

We denote the angle between the rotational axis and the magnetic
dipole moment, or the magnetic inclination, as $\theta_{\rm m}$ and
the angle between the rotational axis and the line of sight as
$\theta_{\rm k}$ (both at their closest approach). If
$\theta_{\rm m}< \theta_{\rm k}$, i.e. the
line-of-sight is outside the magnetic inclination cone, the 
maximum variation of the PA is
\begin{equation}
  \label{eq:27}
  \pm \Delta \mr{PA}_{\rm max} = \pm\,
\mr{arctan} (\mr{tan}\,\theta_{\rm m}/\sin \theta_{\rm k}),
\end{equation}
which is close to $\pm \theta_{\rm m}$ if $\theta_{\rm k}$ is of order
$\sim$$1\,$rad. On the other hand, if 
$\theta_{\rm m}>\theta_{\rm k}$, i.e. the
line-of-sight is inside the magnetic inclination cone, then the PA can change
up to 180$^{\rm o}$ from burst to burst. The PAs of the repeating FRB
121102 varied by $\pm \Delta \mr{PA}_{\rm max}\sim\pm
20^{\rm o}$ for 26 bursts detected over a 7 month period
\citep{2018Natur.553..182M, 2018arXiv180404101G}. This suggests 
that (1) the magnetic inclination angle for the neutron
star associated with 121102 is $\theta_{\rm m}\lesssim20^{\rm
  o}$ and (2) the observer is outside the magnetic inclination cone: 
$\theta_{\rm m}<\theta_{\rm k}$.

By eq. (\ref{eq:Rfo2}), the condition $R_{\rm fo}\gtrsim 10R_{\rm ns}$
means that $B_{0,15}/P\gtrsim 0.04\, 
L_{41}^{1/2}\mc{M}^{-1} R_{\rm fo}/R_{\rm B}$
or the dipole spin-down time $t_{\rm 
  sd}\lesssim 10^4\,\mr{yr}\, L_{41}^{-1}(\mc{M} R_{\rm
  B}/R_{\rm fo})^2$. We see that this condition can be easily satisfied by
young neutron stars, especially if the multiplicity 
factor\footnote{
In the magnetar model, the
magnetic field anchored on the active stellar crust leads to a twisted
external magnetosphere with strong persistent currents
\citep{2002ApJ...574..332T}. For a large global twist angle of
$\sim$$1\,$rad near the surface, the current is given by
$j = |\nabla\times \vec{B}|c/4\pi\sim Bc/4\pi r$, so the
multiplicity factor $\mc{M}(r\sim R_{\rm ns}) \gtrsim jP/B \sim
2.4\times10^3 P$, so $\mc{M}/P\gg 1$ near the surface. At radius $r\gg
R_{\rm ns}$, the current due to magnetospheric twist drops as
$j\propto r^{-4}$ or faster (if the twist does not extend to large
radii), so the multiplicity factor decreases as $\mc{M}\propto r^{-1}$
or faster.} $\mc{M}\gg 1$ near $R_{\rm fo}$. On the other hand, by
eq. (\ref{eq:Rfo3}), the condition 
$R_{\rm L}\gtrsim 10R_{\rm fo}$ means that
\begin{equation}
  \label{eq:18}
  {B_{0,15}\over P^2}\lesssim
2 {L_{41}^{1/2} (R_{\rm fo}/R_{\rm B})\over \mc{M}}\lesssim 2L_{41}^{1/2},
\end{equation}
where we have used $\mc{M}\gtrsim 1$ and
$R_{\rm B}(R_{\rm fo})\gtrsim R_{\rm fo}$ for the second
inequality. Thus, the dipole spin-down 
luminosity $L_{\rm sd}\lesssim 10^{38}L_{41}\,\mr{erg\,s^{-1}}$, and the
neutron star is rotating slowly:
\begin{equation}
  \label{eq:26}
  P\gtrsim
(0.7\,\mr{sec})\,B_{0,15}^{1/2}L_{41}^{-1/4}.
\end{equation}
This constrains the dipole spin-down time to be $t_{\rm
  sd}\gtrsim (15\mr{\,yr})\, B_{0,15}^{-1}L_{41}^{-1/2}$, which is
similar to the age constraint from the non-detection of the time
derivative of DM due to the expanding supernova remnant
\citep{2016ApJ...824L..32P}.
We also note that the PA stays nearly constant over the
duration of each burst \citep[within error 
$\delta \mr{PA}\lesssim 0.1\,$rad,][]{2018Natur.553..182M}. This can
be understood if the neutron star is a slow rotator $\Omega
\Delta t\lesssim (\delta \mr{PA}/\Delta \mr{PA}_{\rm max})\pi/2$ or $P>
0.01(\Delta t/1\mr{\,ms})\,$sec, where $\Delta t$ is the burst
duration.  This is a weaker constraint than eq. (\ref{eq:26}).

We conclude that the polarization observations for FRB 121102
suggests: (1) the emitting region is not far from the neutron star's
surface (since $R_{\rm s}\ll R_{\rm fo}\ll R_{\rm L}$); and (2) 
the neutron star is a slow rotator with period $P\gtrsim
(1\mr{\,sec})B_{0,15}^{1/2}$ (ignoring the weak dependence of
$L_{41}^{-1/4}$). These two pieces of information have 
interesting implications about the radiation mechanism.

The coherent curvature
radiation (or the antenna mechanism) requires high charge density to
produce the high FRB luminosities, and strong magnetic field to
maintain the coherence of charge bunches, which means it can only
operate within a few $R_{\rm ns}$ of the neutron star surface
\citep{2017MNRAS.468.2726K, 2018MNRAS.477.2470L}.  This model is consistent with the polarization constraints discussed in this paper. In the ``cosmic
comb'' model of \citet{2017ApJ...836L..32Z, 2018ApJ...854L..21Z}, the
EM waves are produced near the light cylinder of the neutron star and
the PA variations among repeating bursts from the same progenitor should be
much larger than $\pm20^{\rm o}$, because the magnetic field direction
changes due to stellar rotation and variations of source locations
$R_{\rm s}$. Most of the proposed maser mechanisms for FRBs in the
literature operate beyond the light cylinder \citep[and hence $R_{\rm
  s}>R_{\rm fo}$, e.g. ][]{2014MNRAS.442L...9L, 2017MNRAS.465L..30G,
  2017ApJ...842...34W, 2017ApJ...843L..26B}, where the magnetic field
configuration is far from a regular dipole field and its direction is
likely to vary by a large amount from burst to burst. Thus, these
models may require fine tuning to explain the observed 100\% linear
polarization and small variations in PA for the bursts from FRB 121102.

\section{Summary}
We have proposed a model to explain the recent observations of nearly 100\%
linear polarization and small variations of the PA for dozens of bursts
detected from FRB 121102 over 7 months (in the host-galaxy comoving
frame).

In this model, the emission is 
generated inside the magnetosphere of a strongly magnetized neutron
star, with a significant fraction of the power initially in the form of
X-mode (with the electric vector perpendicular to the
$\vec{k}${--}$\vec{B}$ plane). As the radio 
waves propagate outwards through the neutron star magnetosphere, the
electric vector adiabatically rotates and stays perpendicular to the local
$\vec{k}${--}$\vec{B}$ plane. At sufficiently large distances from the
neutron star surface, this ``adiabatic walking'' behavior stops (and the PA is
frozen), when the plasma density is too small to be able to force the
electric vector to follow the rotation of the local magnetic
field along the propagation. The wave at the exit point is nearly
100\% linearly polarized and the PA seen by an observer at infinity is
determined by the local magnetic field orientation at this point. 
The small range of PA variations from FRB 121102 is naturally
explained, when the following two conditions are satisfied: (1) the
exit point (or freeze-out radius $R_{\rm fo}$) is far away 
from the neutron star surface and well inside the light 
cylinder, i.e. $R_{\rm ns}\ll R_{\rm fo}\ll R_{\rm L}$; (2) the
emission radius $R_{\rm s}$ is much below the exit point $R_{\rm s}\ll R_{\rm
  fo}$. Under these two conditions, the magnetic field at the exit point is
expected to be nearly dipolar and the wave vector (or the observer's
line-of-sight) is nearly in the radial direction, so the direction of the wave
electric field at the exit point is parallel to $\vec{k}\times\vec{B}\parallel
\vec{k}\times\vec{m}$, where $\vec{m}$ is the neutron star's magnetic
dipole moment. We see that the PA is always perpendicular to
$\vec{m}$ projected in the
plane of the sky, independent of the (highly uncertain) magnetic field
configuration near the source region.

Since $R_{\rm s}\ll R_{\rm fo}\ll R_{\rm L}$, the emission radius is
near the surface of the neutron star. This lends support to the 
coherent curvature radiation mechanism which requires high charge
density $n\gtrsim 10^{17}\rm\,cm^{-3}$ to generate the high luminosity
and strong magnetic field $B\gtrsim 10^{14}\,$G to 
maintain the coherence of charge bunches
\citep{2017MNRAS.468.2726K}. Many other models in the literature are
based on strong collisions of magnetized gas flows near 
or beyond the light cylinder \citep{2014MNRAS.442L...9L, 2017MNRAS.465L..30G,
  2017ApJ...842...34W, 2018ApJ...854L..21Z}, where the magnetic field orientation is
irregular and the PA may vary randomly over a much larger range than
observed from FRB 121102. 



Under the model proposed in this paper, we use the PA variations from
FRB 121102 to constrain the angle between the rotation and magnetic
axes (or magnetic inclination) of the underlying neutron star to be
$\theta_{\rm m}\lesssim 20^{\rm o}$. Moreover, the
requirement $R_{\rm ns}\ll R_{\rm fo}\ll R_{\rm L}$ constrains the
rotation period of the neutron star to be $P\gtrsim
(1\,\mr{sec})\,B_{0,15}^{1/2}$, where $B_0= 10^{15}B_{0,15}\,$G is the
surface dipole magnetic field strength.

We predict that the
burst-to-burst variation of PAs from FRB 121102 (at the same
frequency) should follow the rotational 
period of the underlying neutron star (but the burst
  occurrence is not necessarily periodic). In the future, when more bursts
from the repeater are detected with polarization measurements, it
should be possible to measure the rotation period, which will provide
crucial support for the neutron star nature of FRB progenitors. Other
repeating FRBs may have a different range of PA variation, depending on the
magnetic inclination $\theta_{\rm m}$ and the observer's
viewing angle with respect to the rotational axis $\theta_{\rm k}$, but the
periodic modulation is still controlled by stellar rotation. The 
detailed PA variation pattern may be used to measure (or constrain)
$\theta_{\rm m}$ as well as $\theta_{\rm k}$.

Another prediction of our model is that, for each
burst, the PA is weakly frequency-dependent. This is because the
freeze-out radius (and hence the local magnetic field orientation) is
frequency-dependent, provided the intrinsic spectrum $\nu L_{\nu}$ is
not flat. We show that the difference between the PAs measured at
$\nu$ and $2\nu$ may be of order $0.1^{\rm o}$ to $1^{\rm o}$,
although the exact dependence is highly uncertain (depending on the
intrinsic FRB spectrum and the detailed magnetospheric field
configuration).

With forthcoming telescopes such as UTMOST \citep{2017MNRAS.468.3746C},
Apertif \citep{2017arXiv170906104M}, FAST \citep{2011IJMPD..20..989N},
ASKAP \citep{2017ApJ...841L..12B}, CHIME 
\citep{2018arXiv180311235T}, the
FRB sample size is expected to grow by a factor of 10{--}100 and many
more repeaters may be found and localized. Our predictions could be
tested given a sufficiently large sample of bursts with polarization
measurements.

Finally, we point out a caveat to our simple model. It is possible
that the magnetosphere has significant global twist at radius $r\gg
R_{\rm ns}$ \citep[e.g.][]{2002ApJ...574..332T}. In this case, the
magnetic field configuration near the exit point is no
longer dipolar (e.g. there may be a significant toroidal component),
so the PA variations for repeating bursts from the same object will
not follow our predictions, which are based on the dipole field assumption. If
this is the case, more sophisticated, time-dependent modeling of the
twisted neutron star magnetosphere \citep[e.g.][]{2013ApJ...774...92P,
  2017ApJ...844..133C} may be needed. Nevertheless, the
  condition for adiabatic walking (eq. \ref{eq:adiabatic-walking}) and
  the location of the freeze-out radius (eq. \ref{eq:Rfo2}) presented
  in this paper would still be useful for probing the large scale
  magnetic field configuration of neutron star magnetospheres.

\section{acknowledgments}
We thank Emily Petroff and Jason Hessels for useful discussions. WL
was supported by the David and Ellen Lee Fellowship at Caltech.
RN was supported in part by NSF grant AST-1312651 and by the Black Hole
Initiative at Harvard University, which is funded by a grant from
the John Templeton Foundation.




\appendix{}
\section{Analytical solution to the linear wave equation}

In this Appendix, we provide an exact solution of the 1-D linear wave
equation (eq. \ref{eq:waveEq1}) along the $z$ direction,
\begin{equation}
  \label{eq:180}
  {c^2\over \omega^2}{\d^2\over \d z^2}(E_x\hat{x}+E_y\hat{y}) +
  \vec{E}_w= -{4\pi i\over\omega} \vec{j}.
\end{equation}
From eq. (\ref{eq:current}), the amplitude of the current density is
\begin{equation}
  \label{eq:300}
  {4\pi i\over \omega} \vec{j} \approx -{\omega_p^2\over
      \omega^2} (\vec{E}_w\cdot \hat{B})\hat{B} + {i\omega_p^2 \over
     \mc{M} \omega_B \omega}\vec{E}_w\times\hat{B},
\end{equation}
where we have ignored the high-order small term
$\mc{O}(\omega_p^2/\omega_B^2)$, see eq. (\ref{eq:24}). We consider the
simple case of a magnetic 
field that is purely in the x-y plane and whose direction rotates
uniformly as a function of $z$,
\begin{equation}
  \label{eq:3}
  \begin{split}
  \vec{B} = B\hat{B} = B(\cos k_Bz\, \hat{x} + \sin k_Bz\,\hat{y}),
\end{split}
\end{equation}
where the magnitude $B$ is independent of $z$. We assume $k_{\rm B}\equiv
R_{\rm B}^{-1}\ll \omega/c$ so that the magnetic field rotates slowly
compared to the wave phase (we are in the WKB regime). We further assume that the
plasma number density $\omega_p$ is independent of $z$. A more general
case with magnetic field at arbitrary inclination with respect to the
z-axis and variable plasma density is considered numerically in
Appendix B.

We define two unit vectors parallel and perpendicular to
the magnetic field in the x-y plane,
\begin{equation}
  \label{eq:4}
  \hat{n}_\para =\hat{B},\ \ 
\hat{n}_\perp = -\sin k_B z\,\hat{x} + \cos k_B z\, \hat{y},
\end{equation}
such that these vectors rotate with the magnetic field,
and we decompose the wave electric field amplitude as
\begin{equation}
  \label{eq:9}
  \vec{E}_w = (E_\para \hat{n}_\para + E_\perp \hat{n}_\perp +
  E_z\hat{z})\mr{e}^{ikz}. 
\end{equation}
The z-component of eq. (\ref{eq:180}) has no derivatives and it gives
\begin{equation}
  \label{eq:23}
  E_z = {i\omega_p^2 \over \mc{M}\omega\omega_B} E_\perp.
\end{equation}
Inside the neutron star's magnetosphere, we have $\omega_p^2
/\mc{M}\omega\omega_B \simeq 4\pi n_{\rm GJ}qc/\omega B\simeq 2/P\nu = 2\times
10^{-9}P^{-1}\nu_9^{-1}$, where $P$ is the rotational period (in sec)
and $\nu_9 = \nu/\,$GHz is the wave frequency. Thus, the z-component
of the wave electric field can be ignored for our problem.

We assume that $E_\para$ and $E_\perp$ are independent of $z$
  and look for rotating eigenmodes each with an eigenvalue $k$ that
  serves the role of a spatial wave-vector. Note that, under this
  approach, we generally have $k\neq \omega/c$. The imaginary part of
  $k$ corresponds to exponential damping or growth, and the real part
  gives the spatial wavelength. The solution to a given physical problem
  is a superposition of the eigenmodes (each propagating
  independently) that satisfies the given boundary conditions at
  $z=0$.

For convenience, below we set $c=1$. The 
components of the wave equation (\ref{eq:180}) in the x-y plane are
\begin{equation}
  \label{eq:190}
  \begin{split}
      {\d^2 \over \d z^2} \left[(E_\para
\hat{n}_\para + E_\perp \hat{n}_\perp) \mr{e}^{ikz} \right] =
  \left[(\omega_p^2 - \omega^2) E_\para\hat{n}_\para
  -\omega^2 E_\perp\hat{n}_\perp\right] \mr{e}^{ikz}.
  \end{split}
\end{equation}
Apart from the $\mr{e}^{ikz}$ factor, the LHS of this equation is $-(k^2 +
k_B^2)(E_\para \hat{n}_\para + 
E_\perp \hat{n}_\perp) + 2ikk_B (E_\para\hat{n}_\perp - E_\perp
\hat{n}_\para)$, so we obtain two equations for the parallel and
perpendicular components:
\begin{equation}
  \label{eq:7}
  \begin{split}
(k^2+k_B^2 + \omega_p^2 - \omega^2) E_\para + 2ikk_BE_\perp & = 0,\\
(k^2+k_B^2 - \omega^2)E_\perp - 2ikk_B E_\para & = 0.
\end{split}
\end{equation}
Any non-trivial solution of the wave equation requires
the determinant of the above linear equations to be zero, so we obtain the following quadratic equation for
$k^2$:
\begin{equation}
  \label{eq:11}
  k^4 - (2\omega^2 - \omega_p^2 + 2k_B^2)k^2 - k_B^2(2\omega^2 -
  \omega_p^2) + \omega^2 (\omega^2 -
  \omega_p^2) = 0.
\end{equation}
The solutions for $k^2$ are real when $\Delta = (2\omega^2
- \omega_p^2 + 2k_B^2)^2 - 4 [-k_B^2(2\omega^2 -
  \omega_p^2) + \omega^2 (\omega^2 -
\omega_p^2)] =\omega_p^4 -8k_B^2\omega_p^2 +
16k_B^2\omega^2\geq 0$. The two branches of solutions are
\begin{equation}
  \label{eq:22}
  k^2 = \omega^2 + k_B^2 - {\omega_p^2\over 2} \pm {\omega_p^2\over 2}\sqrt{1 -
  {8k_B^2\over \omega_p^2} + {16k_B^2\omega^2\over\omega_p^4}},
\end{equation}
which correspond to two eigenmodes of different polarization states
propagating towards $+z$ (and there are two more propagating in the
opposite direction).

In the following, we discuss two cases: (1) high plasma density
$\omega_p\gg \omega$ (near the neutron star surface); (2) low plasma
density $\omega_p\ll \omega$ (far from the neutron star surface).

\vspace{0.3cm}

\noindent\textbf{Case (1):} $\omega_p\gg \omega\gg
k_{\rm B}$; we have $\Delta >0$. 
The eigenvalues of the two modes are given by
\begin{equation}
  \label{eq:13}
  k_+^2 \approx \omega^2 -k_B^2+ {4k_B^2\omega^2\over \omega_p^2},\ \
k_-^2 \approx \omega^2 -\omega_p^2 + 3k_B^2 -
{4k_B^2\omega^2\over\omega_p^2}. 
\end{equation}
In this case, the $k_-^2$($<0$) branch corresponds to the
O-mode and doesn't propagate along the z-direction\footnote{
This exponentially decaying solution is due to the specific geometry
assumed, i.e. the magnetic field is perpendicular to the allowed
direction of wave propagation. When $\omega_p\gg \omega$, the O-mode waves
can only propagate along the magnetic field (see \S 2.3).} (because
$\omega_p\gg \omega$). The $k_+$
solution corresponds to the X-mode, and it satisfies
\begin{equation}
  \label{eq:14}
 \left| {E_\para\over E_\perp}\right| = {k_+^2 - \omega^2 + k_B^2\over 2k_+k_B}
  \approx {2k_B \omega \over \omega_p^2}\ll 1.
\end{equation}
This shows that the electric vector of this eigenmode 
is nearly perpendicular to the local
magnetic field. In contrast, the electric vector of the (non-propagating) $k_-$
eigenmode is nearly parallel to the field.

Consider a wave with frequency $\omega$ that propagates towards $+z$
and is initialized at $z=0$ with its electric field lying in the
$xy$-plane, perpendicular to the local $\vec{B}$. Any electric field
in this plane can be written as a linear sum of the two eigenmodes of
the problem. In the present case, the sum will be dominated by the
$k_+$ mode, with only a small amount (of order $\omega
k_B/\omega_p^2$) of the $k_-$ mode. With increasing $z$, the $k_-$
mode will decay (because it is non-propagating), and the wave will
become a pure $k_+$ mode, with its polarization rotating so as to
remain nearly perpendicular to the magnetic field. Thus the wave
adiabatically walks with the rotating field.

\vspace{0.3cm}

\noindent\textbf{Case (2):} $\omega\gg \omega_p\gg
k_{\rm B}$; $\Delta >0$ for this case as well. The square-root term in
solutions (\ref{eq:22}) has different asymptotic behaviors in the
following two sub-regimes. The first regime is when
$\omega_p^4/16k_B^2\omega^2\gg 1$, and we have
\begin{equation}
  \label{eq:17}
k^2 \approx \omega^2 + k_B^2 - {\omega_p^2\over2} \pm
\left({\omega_p^2\over 2} - 2k_B^2 + {4k_B^2\omega^2\over
    \omega_p^2}\right), 
\end{equation}
that is,
\begin{equation}
  \label{eq:20}
  \begin{split}
      k_+^2 & \approx \omega^2 - k_B^2 + {4k_B^2\omega^2\over \omega_p^2}\\
      & \Longrightarrow\
\left| {E_\para\over E_\perp}\right| = \left|{k_+^2 - \omega^2 + k_B^2\over
    2k_+k_B}\right| \approx {2k_B \omega \over \omega_p^2}\ll 1, \\
k_-^2 &\approx \omega^2 - \omega_p^2 + 3k_B^2 - {4k_B^2\omega^2\over
  \omega_p^2}\\
  & \Longrightarrow \left| {E_\perp\over E_\para}\right| = \left|{k_-^2 -
    \omega^2+\omega_p^2 + k_B^2\over
    2k_-k_B}\right| \approx {2k_B\omega \over \omega_p^2}\ll 1.
  \end{split}
\end{equation}
We can see that in this regime, the $k_+$ and $k_-$ solutions
correspond to the X-mode ($E_\perp$ dominates) and O-mode ($E_\para$
dominates) respectively. Both of these are propagating modes, and their electric
vectors rotate adiabatically to keep $\vec{E}$ nearly perpendicular to 
$\vec{B}$ and $\vec{k}$ (X-mode) or in the $\vec{k}$--$\vec{B}$ plane 
(O-mode). As in the previous case, a wave that starts off with its
electric field perpendicular to $\vec{B}$ at $z=0$ will adiabatically
rotate with the field and will remain nearly perpendicular to $\vec{B}$
as it propagates.

The second regime is when $\omega_p^4/16k_B^2\omega^2\ll 1$. Here we find
\begin{equation}
  \label{eq:16}
k^2 \approx \omega^2 + k_B^2 - {\omega_p^2\over 2} \pm 2k_B\omega \left(1 +
  {\omega_p^4\over  32k_B^2\omega^2} - {\omega_p^2\over 4\omega^2}
\right).
\end{equation}
If we retain only the lowest order terms, we have
$k_\pm^2 \approx \omega^2 \pm 2k_B\omega - \omega_p^2/2$ and $k_{\pm}\approx
\omega \pm k_B - \omega_p^2/4\omega$, which means
\begin{equation}
  \label{eq:19}
  \begin{split}
  \left({E_\para\over E_\perp}\right)_{\pm} & = {k_\pm^2 - \omega^2 + k_B^2\over
    2ikk_B} \approx {-\omega_p^2 \pm 4k_B\omega\over 4ikk_B} \\
   &\approx i \left(\mp 1 +
    {\omega_p^2\over 4k_B\omega} \right).
\end{split}
\end{equation}
The two eigenmodes have electric fields
\begin{equation}
  \label{eq:21}
  \begin{split}
      \vec{E}_\pm &= (E_{\para\pm}\hat{n}_\para + E_{\perp\pm} \hat{n}_\perp)
  \mr{e}^{ik_\pm z} \\ 
   & \approx \left[ {i\omega_p^2\over 4k_B\omega} \hat{n}_\para  
     \mp i\hat{n}_\para + \hat{n}_\perp\right] E_{\perp\pm} \mr{e}^{i(\omega\pm k_B -
\omega_p^2/4\omega)z} \\
& = \left[ {i\omega_p^2\over 4k_B\omega} \hat{n}_\para \mr{e}^{\pm i 
   k_B  z}  + (\hat{y} \mp i\hat{x}) \right] E_{\perp\pm}
\mr{e}^{i\omega z (1 - \omega_p^2/4\omega^2)},
  \end{split}
\end{equation}
where $E_{\perp\pm} \approx (1 \pm \omega_p^2/8k_B\omega)/\sqrt{2}$
are determined by normalization $|\vec{E}_\pm|=1$.
Since $\omega_p^2/4k_B\omega\ll 1$, the first term corresponds to a
weak rotation of the electric vector at the rate of $k_B$. 
The second term represents the commonly used left and right circularly 
polarized modes. If we ignore the first term, any linearly polarized wave at
$z=0$ can be decomposed into the superposition of $\vec{E}_+$ and
$\vec{E}_-$ with equal amplitudes. Then, as the wave propagates, it
stays linearly polarized with the PA unchanged. 
However, because of the first term, the PA rotates
by $\sim \omega_p^2 \Delta z/(2\omega)$ when the wave travels
a distance $\Delta z\lesssim k_B^{-1}$.

To summarize, we have shown that, in the linear regime, the wave
electric vector adiabatically walks with the twisting magnetic field
as long as the following condition is satisfied: $\omega_p^2 \gtrsim 
k_B\omega/2$. When the plasma density drops such that $\omega_p^2  
\ll k_B\omega/2$, the electric vector stops corotating with the
plasma magnetic field and asymptotically approaches vacuum-like
propagation. The transition from adiabatic walking to polarization
freeze-out happens when $\omega_p^2 \sim ck_B\omega$ 
(where we have restored the $c$). Writing $k_B = 2\pi/R_B$ and
using $\omega \simeq ck = c/\lambdabar$, the freeze-out occurs when
\begin{equation}
\frac{\omega_p^2}{\omega^2} \gtrsim \frac{\lambdabar}{R_B},
\end{equation}
which agrees with equation (\ref{eq:freezeout1}).

To derive the condition (\ref{eq:adiabatic-walking}) in the main
text, we start with equation (\ref{eq:300}) and consider
the parallel component of the current:
\begin{equation}
j_{\parallel} = \frac{i\omega_p^2}{4\pi\omega} E_{\parallel}.
\end{equation}
The X-mode is dominated by the perpendicular field, so
$E_{\perp}\approx E_{w}$, the total electric field. Equations
(\ref{eq:14}) and (\ref{eq:20}) show that, for both Case (1) and Case
(2) with adiabatic walking, the ratio of $E_\parallel$ to $E_\perp$ is
small and equal to $2ck_B\omega/\omega_p^2$ (again restoring
$c$). Thus, the parallel current can be written as
\begin{equation}
|j_{\parallel}| \approx \frac{ck_B}{2\pi} E_{w}.
\end{equation}  
As argued in the main text, the maximum current that the plasma can
support is $qnc$. Therefore, polarization freeze-out will occur once
we have
\begin{equation}
\frac{ck_B}{2\pi} E_{w} \gtrsim qnc,
\end{equation}
which can be rewritten as the condition
\begin{equation}
\frac{\omega_p^2}{\omega^2} \gtrsim \frac{a_0 \lambdabar}{R_B},
\ \ a_0 \equiv \frac{qE_w}{mc\omega}.
\end{equation}
This agrees with equation (\ref{eq:adiabatic-walking}).

\section{Numerical solution to the linear wave equation}

\begin{figure*}
  \centering
\includegraphics[width = 0.53 \textwidth,
 height=0.28\textheight]{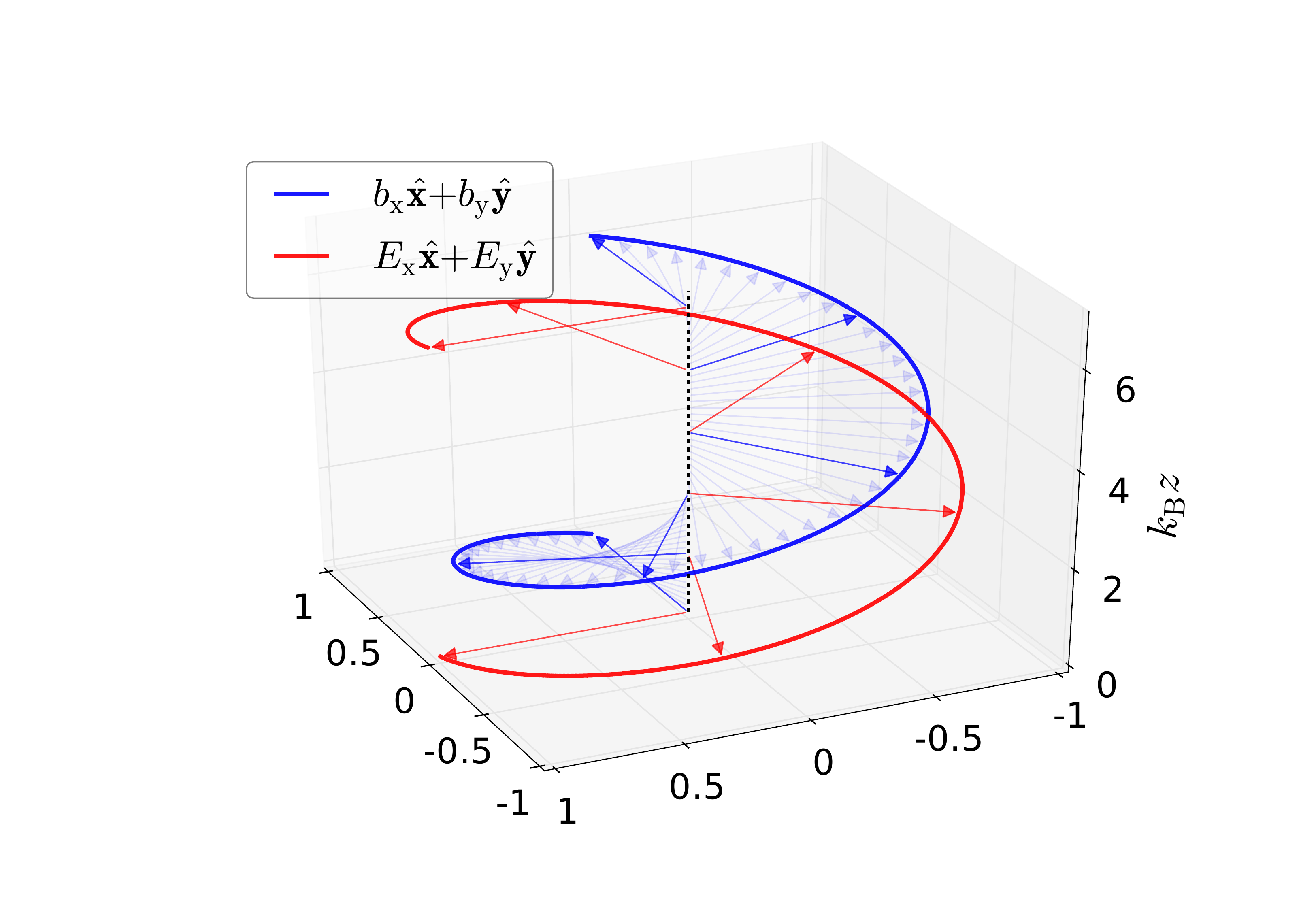}
\includegraphics[width = 0.42 \textwidth,
 height=0.23\textheight]{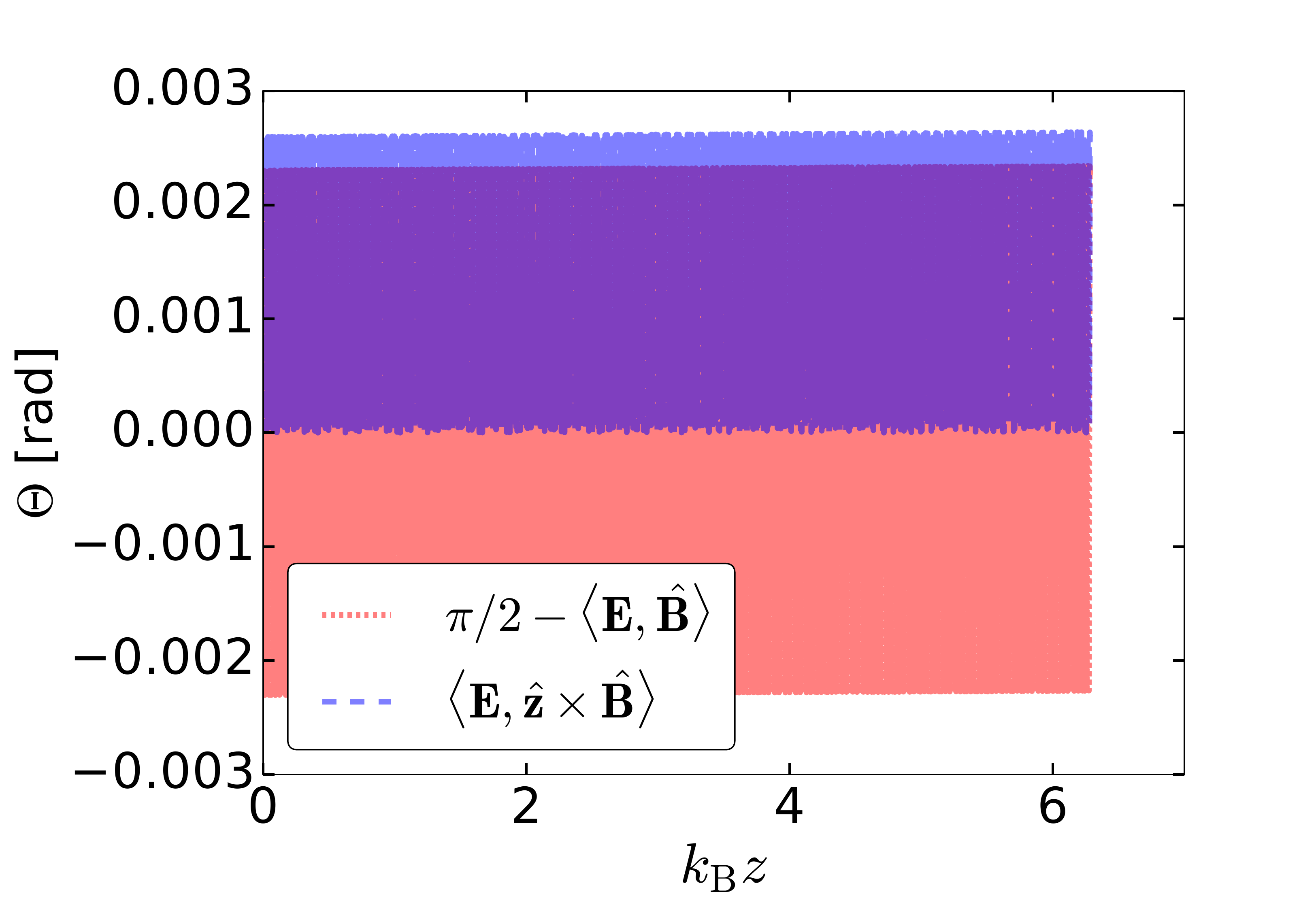}
\caption{Numerical solution to the wave
equation (\ref{eq:100}) with constant $\omega_{\rm p}^2k /\omega^2
k_{\rm B}=1000$, $\theta = 60^{\rm o}$ and boundary conditions: 
$\vec{E}_{\rm w}(z=0) = \hat{y}$ (normalized so that $|E_{\rm w}|=1$)
and $\d \vec{E}_{\rm w}/\d z_1 (z=0) = 0$. \textit{Left Panel:} The
wave electric vector (red arrows) rotates with the turning of static 
magnetic field (blue arrows, only the projection in the x-y plane is
show) in a way such that $\vec{E}_w$ stays parallel to 
$\hat{z}\times\vec{B}$. \textit{Right Panel:} The angle between the
wave electric vector and $\hat{z}\times \hat{B}$ is very small (blue,
which agrees will the analytical estimation in eq. \ref{eq:10}),
and the angle between the wave electric vector and the static magnetic
field is very close to $\pi/2$ (red). Both angles oscillate rapidly on
length scale of $(\omega^2/\omega_{\rm p}^2)k^{-1}= 10^{-3}k_{\rm
  B}^{-1}$.
}\label{fig:const}
\end{figure*}

In this Appendix, we numerically integrate the linear wave equation in one
dimension along the z-axis, for variable plasma density and magnetic
field at arbitrary inclination with respect to the z-axis.

We combine the current in eq. (\ref{eq:current}) and the general wave
equation (\ref{eq:waveEq1}), and obtain
\begin{equation}
    {c^2\over \omega^2}{\d ^2\over \d z^2}(E_{x}\hat{x} +
    E_{y}\hat{y}) + \vec{E}_w= {\omega_p^2\over
      \omega^2} (\vec{E}_w\cdot \hat{B})\hat{B} - {i\omega_p^2 \over
     \mc{M} \omega_B \omega}\vec{E}_w\times\hat{B},
\end{equation}
where we have ignored the high order small current term
$\mc{O}(\omega_p^2/\omega_B^2)$ (see eq. \ref{eq:24}). In
the following, we take a 
different approach from Appendix A and write the wave
amplitude in Cartesian basis as $[E_x(z_1)\hat{x} + E_y(z_1) \hat{y} +
E_z(z_1)\hat{z}]\mr{exp(iz_1)}$, where $z_1 \equiv 
kz$ and $k \equiv \omega/c$ is a constant. Then we obtain a
set of coupled ordinary differential equations 
\begin{equation}
  \label{eq:100}
\left({\d^2 \over\d z_1^2} + 2i
  {\d \over\d z_1} \right)
\left[ {\begin{array}{c}
  E_{\rm x} \\
   E_{\rm y} \\
  \end{array} } \right]
= {\overset\leftrightarrow{A} \over 
\omega^2/\omega_{\rm p}^2 - b_{\rm z}^2}
\cdot
\left[ {\begin{array}{c}
  E_{\rm x} \\
   E_{\rm y} \\
  \end{array} } \right],
\end{equation}
where
\begin{equation}
  \label{eq:1}
\overset\leftrightarrow{A} = \left[ {\begin{array}{cc}
  b_{\rm x}b_{\rm x} & b_{\rm x}b_{\rm y} +\chi \\
   b_{\rm x}b_{\rm y} - \chi
 & b_{\rm y}b_{\rm y} \\
  \end{array} } \right],\, \chi = {i\omega \over \mc{M}\omega_{\rm B}}
({\omega_{\rm p}^2\over \omega^2} - 1)b_{\rm z},
\end{equation}
\begin{equation}
  \label{eq:101}
  E_{\rm z} = {(E_{\rm x} b_{\rm x} + E_{\rm y} b_{\rm y})b_{\rm z} -
{i\omega \over \mc{M}\omega_{\rm B}} (E_{\rm x} b_{\rm y} -
E_{\rm y}b_{\rm x}) \over \omega^2/\omega_{\rm
  p}^2  - b_{\rm z}^2},
\end{equation}
and $b_{\rm x}$, $b_{\rm y}$, $b_{\rm z}$ are the Cartesian components
of the unit vector $\hat{B}$ (along the direction of the static magnetic
field). 

We consider the orientation of the magnetic field rotating as a
function of $z$ at arbitrary pitch angle $\theta$ with respect to the z-axis 
\begin{equation}
  \label{eq:396}
  \hat{B} = \sin\theta\cos k_{\rm B}z \,\hat{x} +
  \sin\theta \sin k_{\rm B}z\, \hat{y} + \cos\theta\, \hat{z},
\end{equation}
where $k_{\rm B} \equiv R_{\rm B}^{-1}\ll k$ (in the WKB regime). If
the plasma properties ($\theta$, $k_{\rm 
B}$, $\omega_p^2/\omega$, $\omega/\mc{M}\omega_B$) are known at every
point $z>0$, we can integrate the 
wave equation (\ref{eq:100}) from the left-hand boundary $z=0$.

We use $\theta = 60^{\rm o}$ and $k_{\rm B}/k =
10^{-4}$ for the two cases presented in this paper. The (X-mode)
boundary conditions are: $\vec{E}_{\rm w}(z=0) = \hat{y}$ (normalized
so that $|E_{\rm w}(z=0)|=1$), $\d \vec{E}_{\rm w}/\d z_1 (z=0)=
0$. We are interested in the situation where $\omega\ll \omega_{\rm
  B}$ and $\mc{M}\geq 1$, so the terms of order $\omega/\mc{M}\omega_{\rm B}$ are
very small. In all cases presented in this paper, we have tested
different choices of $\omega/\mc{M}\omega_{\rm B} = 10^{-3}$
(corresponding to $B\gtrsim 10^6\,$G), $10^{-6}$ and $10^{-9}$, and 
our results are practically the same.

In the realistic
neutron star magnetosphere, the ratio $k_{\rm 
  B}/k$ is much smaller than $10^{-4}$ near the freeze-out radius, and
our calculations (limited by computational cost) can be appropriately
scaled to arbitrarily small $k_{\rm B}/k$. In Fig. \ref{fig:const}, we show
the case with constant $\omega_p^2{k}/\omega^2k_{\rm B} = 1000$, which
corresponds to the propagation below the freeze-out radius given by
eq. (\ref{eq:Rfo1}) in the linear regime. In Fig. \ref{fig:var}, we show the case
with decreasing ratio $\omega_p^2{k}/\omega^2k_{\rm B}=1000 (k_{\rm
  B}z+1)^{-5}$ along the propagation. For
this case, we can see that the transition from adiabatic walking
(rotating electric vector) to vacuum-like propagation (non-rotating
electric vector) occurs at $k_{\rm B}z\simeq 2$ and is quite sharp.

\end{document}